Materials Pushing the Application Limits of Wire Grid Polarizers further into the Deep Ultraviolet Spectral Range

*Thomas Siefke\*, Stefanie Kroker, Kristin Pfeiffer, Oliver Puffky,*

*Kay Dietrich, Daniel Franta, Ivan Ohlídal, Adriana Szeghalmi, Ernst-Bernhard Kley and*

*Andreas Tünnermann*

T. Siefke, K. Pfeiffer, O. Puffky, K. Dietrich, Dr. E.-B. Kley, Prof. Dr. A. Tünnermann

Institute of Applied Physics, Friedrich-Schiller-University Jena, Albert-Einstein-Straße 15, 07745 Jena, Germany

E-mail: Thomas.Siefke@uni-jena.de

Prof. Dr. S. Kroker

Technische Universität Braunschweig, Pockelsstraße 14, 38106 Braunschweig, Germany

Physikalisch-Technische Bundesanstalt, Bundesallee 100, 38116 Braunschweig, Germany

Dr. D. Franta, Prof. Dr. I. Ohlídal

Faculty of Science, Masaryk University, Kotlářská 2, CZ-611 37 Brno, Czech Republic

Wire grid polarizers (WGPs), periodic nano-optical meta-surfaces, are convenient polarizing elements for many optical applications. However, they are still inadequate in the deep ultraviolet spectral range. We show that to achieve high performance ultraviolet WGPs a material with large absolute value of the complex permittivity and extinction coefficient at the wavelength of interest has to be utilized. This requirement is compared to refractive index models considering intraband and interband absorption processes. We elucidate why the extinction ratio of metallic WGPs intrinsically humble in the deep ultraviolet, whereas wide bandgap semiconductors are superior material candidates in this spectral range. To demonstrate this, we present the design, fabrication and optical characterization of a titanium dioxide WGP. At a wavelength of 193 nm an unprecedented extinction ratio of 384 and a transmittance of 10 % is achieved.

# 1    Introduction

Recent progress in functional micro- and nanostructures has opened extensive possibilities for the realization of optical meta-surfaces which are able to shape light in its propagation behavior and all its properties. [1-6] Notably, the preparation of light with a defined polarization state is a key requirement in numerous photonic applications. For example, linear optical polarizers are frequently utilized in lithography,[7,8] industrial vision,[9] microscopy, ellipsometry[10] or astronomic remote sensing systems.[11] All these applications substantially benefit from efficient nano-optical wire grid polarizers.

A wire grid polarizer (WGP) is a grating type meta-surface (see **Figure 1**). The typical operation principle for such elements requires the transmittance of TM polarized light $T_{\text{TM}}$ (TM transversal magnetic - electrical field orthogonal to the ridges) to be much larger than that of TE polarized light $T_{\text{TE}}$ (transversal electric - electrical field parallel to the ridges) to achieve a significant anisotropic filter functionality. Here, the extinction ratio $Er = T_{\text{TM}}/T_{\text{TE}}$ is used to express the suppression of TE polarized light.[14]

WGPs are highly beneficial because of large achievable element sizes (wafer size), compactness (wafer thickness) and large acceptance angles.[12] Furthermore, their nano-optical nature allows an easy integration into other (nano-)optical elements, such as lithography masks, [13] enabling local polarization control. Currently, applications advance towards shorter wavelengths in order to benefit from smaller foci and characteristic electronic transitions, which can be utilized for material analysis. While WGPs are well established in the VIS and IR, suitable ones were not available in the deep ultraviolet (DUV) spectral range until very recently. [14, 16, 37] The lack of applicable DUV WGPs originates from challenging requirements on both structure and material properties.

A structural period of the polarizer has to fulfilling the zero order condition:

$$p < \lambda/(n_{\text{sub}} + sin(\varphi)) \qquad (1)$$

to avoid propagation of diffraction orders greater than the zeroth one.[17] For a normal incidence of light ($\varphi = 0°$) with a wavelength $\lambda$ in the DUV and a fused silica substrate with a refractive index $n_{\text{sub}}$ a period $p$ in the order of 100 nm is necessary. Additionally, an aspect ratio (see Figure 1: ratio of height and ridge width) larger than five is typically required. [14] The simultaneous realization of large aspect ratio and small periods is technologically extremely challenging. Fortunately however, advances in nanotechnology do allow the fabrication of such structures. [15] Pelletier et al. [16] demonstrated aluminum WGPs with a period of 33 nm, which is three times smaller than other WGPs intended for use in the DUV. [14] The measurements and numerical calculations from Pelletier et al. show a rapid decline of the extinction ratio towards smaller wavelengths below about 300 nm rendering these elements as inappropriate for DUV applications. Based on these results we conclude that solely reducing the structural period is not sufficient to fabricate WGPs in the DUV. But additionally, an appropriate material is necessary to achieve high DUV performance WGPs.

In this publication, we begin with a qualitative discussion of the requirements imposed on the complex refractive index for DUV WGP applications. These requirements explain the unsuitability of the commonly used Drude metals, for example aluminum, for application in DUV WGPs. Then, we demonstrate the success of materials with appropriate electronic interband transitions in achieving this same goal. Therefore, titanium dioxide WGPs are designed, fabricated and characterized. The retrieved experimental results are compared to measurement data for several WGPs made of different materials (metals as well as semiconductors) and discussed with respect to the material requirements.

## 2    Material requirements for efficient wire grid polarizers

To understand the demands on a suitable material for WGPs the electric field boundary conditions at the horizontal interfaces are considered.[17]

$$\vec{E}_{TE}^{ridge} = \vec{E}_{TE}^{sur} \qquad \text{for TE} \qquad (2)$$

$$\varepsilon^{ridge}\vec{E}_{TM}^{ridge} = \varepsilon^{sur}\vec{E}_{TM}^{sur} \qquad \text{for TM, where} \qquad (3)$$

$$|\varepsilon|/\varepsilon_0 = |\varepsilon_r| = n^2 + k^2 \qquad (4)$$

Here, $\vec{E}$ is the corresponding complex electric field either in the ridge $\vec{E}^{ridge}$ or in the surrounding medium $\vec{E}^{sur}$, $\varepsilon^{ridge}$ and $\varepsilon^{sur}$ is the complex permittivity of the ridge material or the surrounding medium, respectively (see Figure 1). Furthermore, $\varepsilon_r$ is the complex relative permittivity and $\varepsilon_0$ is the vacuum permittivity. In the following, the surrounding medium is assumed to be vacuum hence: $\varepsilon^{sur} = 1 \cdot \varepsilon_0 = \varepsilon^{vac}$. Note that in the theory of dispersion the quantity complex relative permittivity is called dielectric function. The absolute value of the complex relative permittivity $\varepsilon_r$ is linked to the complex refractive index $\tilde{n} = n + ik$ via Equation 4, where $n$ is the refractive index and $k$ the extinction coefficient. Equation 2 denotes that the electric field is continuous at interfaces. Regarding a grating structure with a period much smaller than the wavelength of the incident light, an incoming TE polarized plane wave is only slightly distorted (see **Figure** 2 a)). Thus, the transmittance $T_{TE}$ becomes small, if the extinction coefficient of the grating material is large.

According to Beer-Lambert law the transmittance may be approximated as:[17]

$$T_{TE} = e^{-\frac{4\pi k_{eff} z}{\lambda}}. \qquad (5)$$

For a given wavelength $\lambda$, and grating height $z$ the transmittance becomes small if the effective extinction coefficient $k_{eff}$ of the structure is large. Were $k_{eff}$ is the extinction coefficient of an assumed homogeneous layer, representing the actual grating structure according to an effective medium approximation.[41]

Regarding TM polarized light three different cases can be distinguished:

**(I)** $|\varepsilon^{\text{ridge}}| \gg |\varepsilon^{\text{vac}}|$ The absolute value of the complex permittivity of the grating material is much larger than that of vacuum. For this configuration, the electric field is much higher in the surrounding vacuum than in the ridges (see Equation 3). Since the extinction coefficient of vacuum is zero, the electric field is only slightly damped (see Figure 2 b)). The transmittance of TM polarized light is therefore very high. This case is the typical operating mode of a WGP.

**(II)** $|\varepsilon^{\text{ridge}}| \gtrsim |\varepsilon^{\text{vac}}|$ The absolute value of the complex permittivity of the grating material approaches that of vacuum. This causes the electric field to be located more inside the absorbing ridges, increasing the damping. Thus, the transmittance of TM polarized light becomes considerably smaller, hence this operation mode is less preferable, if applicable at all.

**(III)** $|\varepsilon^{\text{ridge}}| \ll |\varepsilon^{\text{vac}}|$ The absolute value of the complex permittivity of the grating material is much smaller than that of vacuum. This causes the electric field to be located inside the ridges (see Figure 2 c)). If the ridge material possesses absorption, the electrical field is strongly damped. With the appropriate material and geometry parameters this can lead to a larger transmittance of TE polarized light than of TM polarized light. This phenomenon is known as the inverse polarization effect. [18]

In summary, both the extinction coefficient $k$ (see Equation 5) and the absolute value of the complex permittivity $|\varepsilon|$ (see case I) should be large, in order to achieve DUV WGPs with high extinction ratio and transmittance.

The stated material requirements can now be linked to material models in order to subsume application wavelength ranges. Actually optical properties of materials are typical a superposition of many processes. The two most important electronic processes in the visible and UV spectra contributing to the optical material properties are intraband and interband transitions. Intraband transitions correspond to the electronic conduction by free carriers. This process is especially relevant for conducting materials such as metals or degenerated semiconductors. Intraband transitions typically dominate the permittivity of these materials in the visible and infrared spectra. In contrast, interband transitions typically arise at shorter wavelengths were the energy of the incident photons becomes larger than the bandgap energy, thus a transition of the electron from the valence band into the conduction band can occur. This absorption of photons is related to an increased extinction coefficient. Interband transitions can be either direct or indirect. The latter requires the generation or absorption of a phonon. Hence, the contribution of indirect transitions to the complex refractive index is generally much weaker than that of direct transitions.[19]

To achieve a comprehension of the particular implications on the WGP performance both processes will be discussed separately in the following sections. To account for the simplification in the theoretical description, the actual performance of WGPs made of several different materials is discussed in section 4.

### 2.1 Wire grid polarizers based on intraband transition processes

Intraband absorption can be classically treated as free carriers in the Drude model.[19] Here the relative permittivity $\varepsilon_\mathrm{r}$ at a frequency $\omega$ is given by:

$$\varepsilon_\mathrm{r}(\omega) = 1 - \frac{\omega_\mathrm{p}^2}{(\omega^2 + i\gamma\omega)}, \qquad (6)$$

with a damping factor $\gamma$ and the plasma frequency $\omega_\mathrm{p}$, which depends on the number of free electrons $n_e$ and the electron mass $m_e$ (to account for solid state - electron interactions the effective mass $m_e^*$ may be used instead):

$$\omega_\mathrm{p} = \sqrt{\frac{n_e e^2}{\varepsilon_0 m_e}}. \tag{7}$$

Exemplarily, the refractive index and the absolute complex permittivity are calculated for aluminum and illustrated in **Figure 3.**

This diagram can be divided into four regimes:

(A): For larger wavelengths the extinction coefficient and absolute value of the complex permittivity are large. This corresponds to case (I) ($|\boldsymbol{\varepsilon}^\mathbf{ridge}| \gg |\boldsymbol{\varepsilon}^\mathbf{vac}|$). The material is therefore well suited for WGPs.

(B): The extinction coefficient becomes smaller. This leads to an increase in the transmittance of TE polarized light. Furthermore, the absolute value of the complex permittivity becomes smaller as well. This corresponds to case (II) ($|\boldsymbol{\varepsilon}^\mathbf{ridge}| \gtrsim |\boldsymbol{\varepsilon}^\mathbf{vac}|$). The electric field for TM polarized light penetrates into the ridges, thus the transmittance of TM polarized light diminishes. In this regime the polarizer is still applicable, but the performance is poor.

(C): For further reduced wavelengths the absolute value of the complex permittivity of the ridges is smaller than that of vacuum (see case (III) ($|\boldsymbol{\varepsilon}^\mathbf{ridge}| \ll |\boldsymbol{\varepsilon}^\mathbf{vac}|$)). In principle, an inverse polarization effect can be observed.

(D): For wavelengths below the plasma wavelength, virtually no absorption is present in the material, thus functionality as a WGP cannot be expected.

It is therefore clear, that the optical performance of a WGP based on pure intraband transition fundamentally diminishes towards short wavelengths. A lower application limit arises, solely caused by material properties and located above $\lambda_\mathrm{P}$ according to case C. Hence, materials with small plasma wavelengths are preferable for applications at short wavelengths. Such WGPs offer very broadband application ranges (UV to IR). Since the loss mechanism for the transmittance $T_\mathrm{TE}$ is based on free electrons, TE polarized light is mostly reflected.

## 2.2  Wire grid polarizers based on interband transition processes

As described above, for interband transition processes, the actual shape of the complex refractive index plot as function of the wavelength is determined by the band structure. This gives typically rise to several distinct features. Such features can be described by a Tauc Lorentz oscillator model. [20] The relative permittivity's imaginary part is:

$$Im\{\varepsilon_r\} = \frac{AE_0\Gamma(E-E_g)^2}{(E^2-E_0^2)^2+\Gamma^2 E^2}\frac{1}{E} \quad E > E_g$$

$$Im\{\varepsilon_r\} = 0, \quad\quad\quad\quad\quad E \leq E_g. \tag{8}$$

Where the parameters are $E$ energy of the incident photon (for a better comparability the results are converted to wavelength $\lambda$), band gap energy $E_g$ (equivalent wavelength denoted as: $\lambda_g$), peak transition energy $E_0$ (equivalent wavelength denoted as: $\lambda_0$), peak broadening $\Gamma$ and parameter $A$ determining the strength of the oscillator proportional to density of electrons. [21] By applying the Kramers-Kronig relation the real part of the complex permittivity and consequently the complex refractive index (**Figure 4**) can be achieved.

Similar to the intraband processes different sections can be assigned (see Figure 4):

(A): In the vicinity of the peak transition energy, both $k$ and $|\varepsilon|$ show a maximum. As previously discussed this is the favorable working regime for a WGP. The maximum of the extinction ratio is expected to be approximately between the maxima of both parameters.

(B): At wavelengths below the peak transition wavelength $k$ as well as $|\varepsilon|$ decrease. Hence, the TM transmittance diminishes and that of TE polarized light increases. Thus, the extinction ratio is reduced.

(D): Since the material has virtually no absorption at wavelengths above the bandgap, the transmittance of TE and TM polarized light is large. A functionality as a WGP cannot be expected. The device is transparent.

To summarize, the extinction ratio of a WGP, based on interband transitions (see Figure 4), has an onset at the bandgap energy and a maximum at the peak transition energy. At shorter wavelengths the extinction ratio decreases. The spectral bandwidth is much narrower than that of WGPs based on intraband processes. In addition, because the loss mechanism for the transmittance is the annihilation of the involved photons, unlike for the mentioned intraband processes, TE polarized light is mostly absorbed.

## 3   Experimental realization of a Titanium oxide wire grid polarizer

In order to experimentally verify the application of materials whose complex refractive index is dominated by intraband processes a titanium dioxide WGP was designed, fabricated and characterized. Titanium dioxide is a wide bandgap semiconductor with a bandgap of about 3.0 - 3.2 eV[22]. Hence, a spectral application range in the DUV is expected. Furthermore, it is a very common material for optical coatings[22], photocatalytic[24,25] and light harvesting applications. [26] **Figure 5** shows the complex refractive index of a titanium dioxide thin film. The fabrication was performed by atomic layer deposition (ALD) (Oxford Instruments OpAL) using titanium tetraisopropoxide (TTIP) and oxygen plasma as precursors at a deposition temperature of 100°C. [27] The film was then characterized by ellipsomety (Jobin Yvon UVISEL2 VUV), transmittance and reflectance spectroscopy (McPhersom VUVas 2000). The refractive index (Figure 5) was obtained by fitting the data using the universal dispersion model. [28,29]

In order to achieve a good performance of a WGP, the grating period, height and ridge width have to be optimized (Figure 1). For an incidence angle of 0 ° ± 20°, a wavelength of 200 nm, the refractive index of fused silica[30] substrates as 1.56 a period of 104.5 is chosen according to Equation. 1. In order to determine ridge width and height, extinction ratio and transmittance of the WGP are simulated by means of Rigorous Coupled Wave Analysis (RCWA) [31] with the measured material parameters of titanium dioxide (Figure 5). The parameters are chosen to achieve a transmittance of TM polarized light of 50 % and an extinction ratio larger than 100. **Figure 6** shows that the extinction ratio and the transmittance are contradictory, i.e. by increasing the ridge height and width $Er$ increases while $T_{TM}$ decreases. From the perspective of fabrication, the ridge height and width should be small. Therefore, we chose a height of 150 nm and a width of 26 nm. According to this design **Figure** 7 shows the simulated extinction ratio and transmittance of a titanium WGP in a wavelength range from 150 nm to 500 nm.

For wavelengths larger than 390 nm (equivalent to the bandgap energy) both TE and TM transmittances are almost 100 %. Thus, the polarizer is transparent and no linear polarizing behavior can be observed ($Er \sim 1$). The spectral position of the maximum extinction ratio at 270 nm is between the maxima extinction coefficient at 258 nm and of the absolute complex permittivity at 290nm (see Figure 5). Since $|\varepsilon|$ is large, the transmittance remains high. Below this wavelength the extinction ratio drops again, but maintains a reasonably large value until approximately 190 nm. Since $|\varepsilon|$ becomes smaller towards shorter wavelengths, the transmittance of TM polarized light decreases as well. From this simulation we expect an application wavelength range from circa 190 nm to 290 nm.

According to the aforementioned design, a WGP is fabricated by a self-aligned double patterning (SADP) process. [32] The optical performance of the fabricated polarizer is then measured by means of a Perkin Elmer Lambda 950 spectrometer equipped with a Glan-Taylor polarizer in the wavelength range from 230 nm to 500 nm and by a laser setup at a wavelength of 193 nm respectively. The measurement results are presented in **Figure** 8**.** The spectral peak position of the extinction ratio is at about 244 nm close to the maximum of the extinction coefficient (see Figure 5). The actual achieved extinction ratio at this wavelength is 834 and a TM transmittance of 15%. At 193 nm wavelength, the achieved extinction ratio is 384 and the transmittance is determined to be 10 %. The measured values differ from the expected ones due to deviation of shape and refractive index from the design values.[32-34] The dip of the transmittance $T_{TM}$ can be attributed to resonance effects caused a grating super-structure introduced by the SADP process in the transparent spectral region.[32]

## 4  Comparison of different wire grid polarizer materials

Material selection is crucial for the performance of a wire grid polarizer (WGP). Therefore, we compare the application wavelength ranges of WGPs fabricated from several materials,[14,35-38] including the titanium dioxide shown above. Using real data, instead of material models, accounts for influences from current fabrication technology and for simplifications in the discussed models (see **Figure 9**).

To enable the comparison of the lower application wavelength limit between these different materials, an extinction ratio larger than 100 is defined as a threshold (horizontal dashed line in Figure 8). Below this value, the performance is regarded as not applicable. Naturally, this threshold depends strongly on the actual application, however a suppression of the erroneous influences of the undesired polarization direction by two orders of magnitude appears reasonable. For simplicity the transmittance of TM polarized light is disregarded, although it is of course an important design parameter.

To begin, the metals aluminum and iridium are discussed. The application limits of aluminum and iridium are 310 and 280 nm, respectively (see Fig. 8). This wavelength difference relates to the different plasma wavelengths of 79 nm and 45 nm (Equation 7). Corresponding to the previous discussions, materials with lower plasma wavelengths enable intraband absorption based wire grid polarizers with lower application wavelengths. Since iridium is one of the materials with the lowest plasma wavelengths (the lowest value is found for cobalt at 39 nm) [39] intraband absorption based polarizers with an application limit at a wavelength far below that of iridium cannot be expected. Thus, the application range of the two compared metallic WGPs is limited to wavelengths ≳280 nm.

The plasma wavelength of tungsten is 54 nm. At first glance, this appears contradictory to the fact that the application limit is found at considerably shorter wavelengths (210 nm) than that of iridium. The explanation for this apparent inconsistency can be found in the band structure of tungsten: Several strong transitions are located in wavelength range below 390 nm[39]. Hence, the complex refractive index of tungsten in this spectral region is dominated by interband transitions enabling a performance of WGPs extended to shorter wavelengths.

The presence of interband transitions in the DUV spectral region explains why semiconductors are increasingly in the spotlight for the realization of nano-optical WGPs. Therefore, in this section we include the behavior of silicon,[38] chromium oxide[6] and our $TiO_2$ WGPs into the discussion. The titanium dioxide WGP shows a strong extinction ratio peak (see Figure 8). This can be attributed to several pronounced interband transitions of $TiO_2$ at wavelengths between 150 and 270 nm[40]. A similar behavior can be found for silicon[38]. The application wavelengths of WGPs range from 190 nm to 290 nm ($TiO_2$) and from 360 nm to 420 nm (Si), respectively. At 248 nm wavelength the extinction ratio of the titanium dioxide WGP ($Er$=774) is much larger than that of tungsten WGPs ($Er$=270). Both, aluminum with an extinction ratio of about 10 and iridium (Er~30) are barely applicable at this wavelength. At 193 nm the transmittance of the titanium dioxide WGP ($T_{TM}$=10 %) is smaller than that of the chromium oxide ($T_{TM}$=18.6 %) and of the tungsten WGPs ($T_{TM}$=44 %). However, its extinction ratio of 384 is substantially better than that of the chromium oxide ($Er$=138) (data only at 193 nm available) and that of the tungsten WGP ($Er$=22). These findings illustrate that interband transitions, particularly in wide bandgap semiconductors, can be utilized to extend the lower application limit of WGPs towards shorter wavelengths in the DUV. Particularly, at wavelengths below about 280 nm nano-optical WGPs based on interband transitions are capable to outperform conventional elements based on intraband transitions. By carefully choosing the meta-surface structure parameters the polarizer's transmittance of TM polarized light and extinction ratio can be specifically balanced to optimally address the requirements of the respective application.

## 5  Conclusion

In this contribution we showed that for efficient meta-surface wire grid polarizers (WGPs), a grating material with high extinction coefficient and simultaneously large absolute value of the complex permittivity has to be used. In consequence, WGPs based on intraband transitions (metals) fail in the deep ultraviolet wavelength range. By utilizing materials whose absorption is based on interband transition processes (especially semiconductors) a superior polarizing performance at much shorter wavelengths can be achieved. This was experimentally verified for a titanium dioxide WGP. An unprecedented extinction ratio of 384 at a wavelength of 193 nm was obtained.

ACKNOWLEDGMENT

The authors would like to thank Helmut Bernitzki for the transmittance measurements of the WGPs at 193 nm and Alexia Stollmann for proof reading.

This research was financially supported by the German Ministry of Education and Science (project NanoInt 13N13021) and the German Science Foundation (Emmy Noether Program SZ 253/1-1 and IRTG 2101).

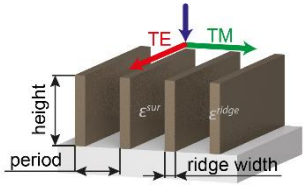

Figure 1: Schematic representation of a wire grid polarizer. The blue arrow indicates the direction of incident light.

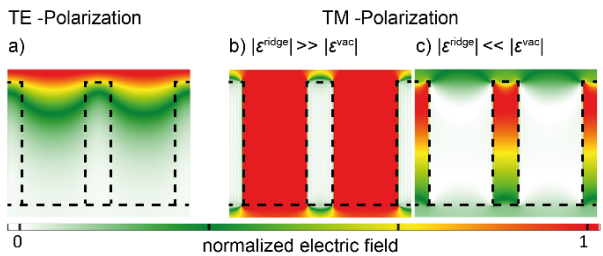

Figure 2: Schematic electric field distribution in a WGP. The dashed lines display the geometry. a) The electric field is exponentially damped for TE - polarization. b) The electric field is located between the ridges, almost no damping occurs for TM-polarization. c) The electric field is located inside the ridges and therefore strongly damped.

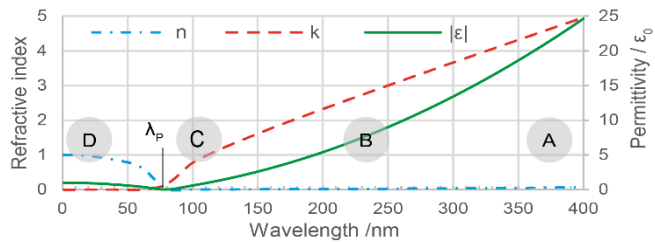

Figure 3: Complex refractive index $\tilde{n} = n + ik$ and $|\varepsilon|$ calculated from Equation 6 for aluminum[16].

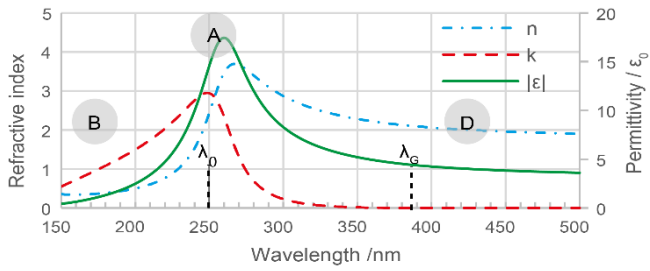

Figure 4: Complex refractive index and absolute complex permittivity calculated by a generic Tauc Lorentz model for a typical wide bandgap semiconductor. $E_g$ =3.2 eV; A = 90 eV; $E_0$ =4.8 eV; Γ=0.6 eV

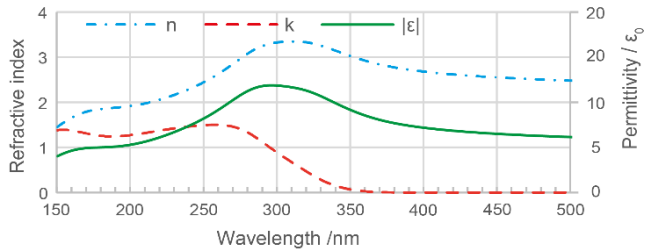

Figure 5: Refractive index (*n*), extinction coefficient (*k*) and absolute value of the complex permittivity |*ε*| of $TiO_2$, deposited by ALD and measured by VUV-ellipsometry and spectroscopy.

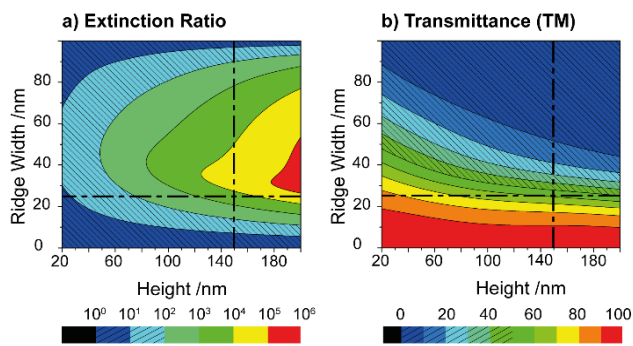

Figure 6: Simulated extinction ratio a) and transmittance of TM polarized light b) dependent on ridge height and width. According to the demanded performance areas with $T_{TM}$ <50% and *Er*<100 are hatched. The points where the ridge width is 26 nm and the ridge height is 150 nm

are marked denoting the final design. The simulation was performed at $\lambda = 248$ nm, $p = 104.5$ nm and $\varphi = 0°$.

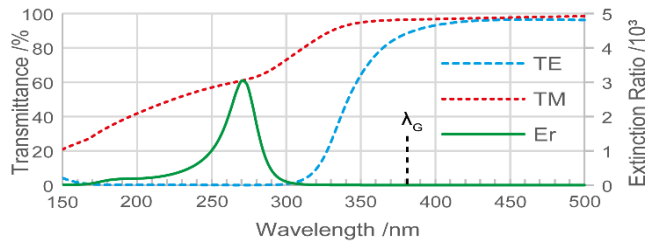

Figure 7: Simulated extinction ratio and transmittance of a titanium dioxide WGP according to the design in Figure 6 between 150 nm and 500 nm. The extinction ratio shows a peak at about 270 nm and remains larger than 100 for wavelength down to about 190 nm.

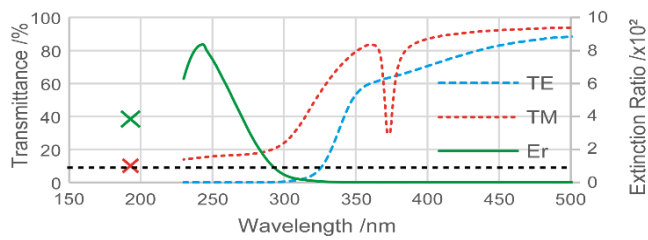

Figure 8: Measured transmittances and extinction ratio of the fabricated titanium dioxide WGP. The red and green crosses mark the measurements at 193 nm of TM transmittance and extinction ratio, respectively. Note that the scale of the extinction ratio is magnified compared to Figure 7.

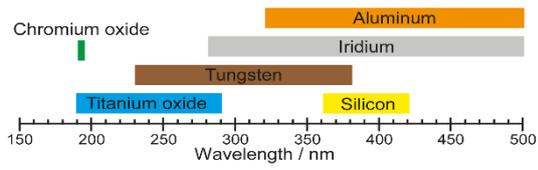

Figure 9: Comparison of application wavelength ranges of several WGP consisting of aluminum[35], iridium[36], tungsten[14], chromium oxide[37], silicon[38] and the here presented TiO$_2$.